\documentclass[
twocolumn, portrait]{aastex631}
\usepackage{rotating}
\usepackage{booktabs}
\usepackage{graphicx}
\usepackage{amsmath,mathtools}
\usepackage{tikz}
\usepackage{CJKutf8} 

\newcommand\omc{$\omega$\,Cen}

\newcommand\deltaone{$\rm{\Delta_{F275W,F814W}}$}
\newcommand\deltatwo{$\rm{\Delta_{C_{F275W,F336W,F435W}}}$}
\newcommand\deltay{$\Delta Y$}

\begin{document}
\begin{CJK*}{UTF8}{gbsn}
\title{oMEGACat V: Helium Enrichment in $\omega$ Centauri as a Function of Metallicity}

\correspondingauthor{Callie Clontz}
\email{clontz@mpia.de}

\author[0009-0005-8057-0031]{C. Clontz}
\affiliation{Department of Physics and Astronomy, University of Utah, Salt Lake City, UT 84112, USA}
\affiliation{Max-Planck-Institut für Astronomie, Königstuhl 17, D-69117 Heidelberg, Germany}

\author[0000-0003-0248-5470]{A. C. Seth}
\affiliation{Department of Physics and Astronomy, University of Utah, Salt Lake City, UT 84112, USA}

\author[0000-0003-2512-6892]{Z. Wang (王梓先)}
\affiliation{Department of Physics and Astronomy, University of Utah, Salt Lake City, UT 84112, USA}

\author[0000-0001-8052-969X]{S. O. Souza}
\affiliation{Max-Planck-Institut für Astronomie, Königstuhl 17, D-69117 Heidelberg, Germany}

\author[0000-0002-5844-4443]{M. H\"aberle}
\affiliation{Max-Planck-Institut für Astronomie, Königstuhl 17, D-69117 Heidelberg, Germany}

\author[0000-0002-2941-4480]{M. S. Nitschai}
\affiliation{Max-Planck-Institut für Astronomie, Königstuhl 17, D-69117 Heidelberg, Germany}

\author[0000-0002-6922-2598]{N. Neumayer}
\affiliation{Max-Planck-Institut für Astronomie, Königstuhl 17, D-69117 Heidelberg, Germany}

\author[0000-0002-7547-6180]{M. Latour}
\affiliation{Institut für Astrophysik und Geophysik, Georg-August-Universität Göttingen, Friedrich-Hund-Platz 1, 37077 Göttingen, Germany}

\author[0000-0002-7547-6180]{A. P. Milone}
\affiliation{Dipartimento di Fisica e Astronomia “Galileo Galilei”, Universita’ di Padova, Vicolo dell’Osservatorio 3, Padova, IT-35122}
\affiliation{Istituto Nazionale di Astrofisica - Osservatorio Astronomico di Padova, Vicolo dell’Osservatorio 5, Padova, IT-35122}

\author[0000-0002-0160-7221]{A. Feldmeier-Krause}
\affiliation{Department of Astrophysics, University of Vienna, T\"urkenschanzstrasse 17, 1180 Wien, Austria}
\affiliation{Max-Planck-Institut für Astronomie, Königstuhl 17, D-69117 Heidelberg, Germany}

\author[0000-0002-6072-6669]{N. Kacharov}
\affiliation{Leibniz Institute for Astrophysics (AIP), An der Sternwarte 16, 14482 Potsdam, Germany}

\author[0000-0001-9673-7397]{M. Libralato}
\affiliation{INAF, Osservatorio Astronomico di Padova, Vicolo dell’Osservatorio 5, Padova,I-35122, Italy}

\author[0000-0003-3858-637X]{A.\ Bellini}
\affiliation{Space Telescope Science Institute, 3700 San Martin drive, Baltimore, MD, 21218, USA}

\author[0000-0003-4546-7731]{G. van de Ven}
\affiliation{Department of Astrophysics, University of Vienna, T\"urkenschanzstrasse 17, 1180 Wien, Austria}

\author[0000-0002-1212-2844]{M. Alfaro-Cuello}
\affiliation{Facultad de Ingenier\'{i}a y Arquitectura, Universidad Central de Chile, Av. Francisco de Aguirre 0405, La Serena, Coquimbo, Chile}

\begin{abstract}
Constraining the helium enhancement in stars is critical for understanding the formation mechanisms of multiple populations in star clusters. However, measuring helium variations for many stars within a cluster remains observationally challenging. We use Hubble Space Telescope photometry combined with MUSE spectroscopic data for over 7,200 red-giant branch stars in \omc\ to measure helium differences between distinct groups of stars as a function of metallicity separating the impact of helium enhancements from other abundance variations on the pseudo-color (chromosome) diagrams. Our results show that stars at all metallicities have subpopulations with significant helium enhancement ($\Delta Y_{min} \gtrsim$ 0.11). We find a rapid increase in helium enhancement from low metallicities ($\rm{[Fe/H] \simeq -2.05}$  to $\rm{[Fe/H] \simeq -1.92})$, with this enhancement leveling out at \deltay\ $= 0.154$ at higher metallicities. The fraction of helium-enhanced stars steadily increases with metallicity ranging from 10\% at $\rm{[Fe/H] \simeq -2.04}$ to over $90\%$ at $\rm{[Fe/H] \simeq -1.04}$. This study is the first to examine helium enhancement across the full range of metallicities in \omc{}, providing new insight into its formation history and additional constraints on enrichment mechanisms.

\end{abstract}


\section{Introduction} \label{sec:intro}

Both photometric and spectroscopic studies of Milky Way globular clusters (GCs) have revealed multiple stellar populations. This includes observations of abundance variations in light elements from high-resolution spectroscopy \citep[e.g.][]{Carretta_2009}, and multiple, sometimes discrete sequences of stars seen in photometry \citep[e.g.][]{Milone_2017a}. These findings have sparked intense interest in GC formation and enrichment histories \citep{Gratton_2019}, but despite this interest, currently there are no models that can fully explain the diversity of populations observed in GCs \citep{Bastian_2018, Milone_2022}.

Constraints on helium enhancements in GCs are critical for understanding the formation of multiple stellar populations. Stars exhibiting high helium abundances indicate that the material from which they formed had undergone high-temperature hydrogen burning. This enhancement in helium is accompanied by abundance variations from several nucleosynthetic processes including the CNO, NeNa, and MgAl cycles \citep{Gratton_2001}. GCs have stars with primordial helium and no evidence for hot hydrogen burning, typically considered the first-generation (1G) stars, as well as stars with enhanced helium and products from hot hydrogen burning, considered the second-generation (2G) stars. 

Helium is difficult to directly measure from spectra due to its sensitivity to the stellar temperature. The atmosphere of stars with $T_\mathrm{eff} > 11\,500$ K is affected by diffusion, causing the helium to sink below the photospheric layers. Thus, any enrichment of helium that can be measured from spectral observations in these hot stars is not reflective of their abundance at formation. Furthermore, for stars with $ \rm{T_{eff}} < 8000$ K the helium lines are chromospheric (He I 10\,830\,$\textup{\AA}$) rather than photospheric, which requires complex models to measure. Stars in the range $8000 < \rm{T_{eff}} < 11\,500$ K, if enriched, will exhibit photospheric helium lines (He I 5\,876\,$\textup{\AA}$) and while these can often be reliably measured \citep{Reddy_2020}, it is observationally expensive to get spectra with high signal-to-noise and sufficient resolution for large samples. 

Helium has been shown to have a significant impact photometrically. In an analysis of 57 globular clusters (including \omc), \cite{Milone_2017a} found that most GCs separate into two distinct sequences on the pseudo-color-color diagrams they refer to as ``chromosome diagrams". They utilize what is known as the ``magic trio" of filters, containing the \textit{Hubble Space Telescope (HST)} photometric bands F275W, F336W, and F435W, which probe the wavelength ranges in which important absorption features \citep{Piotto_2015} lie. This enables parsing of subpopulations with varied abundances from CNO-cycling. They find two sequences in most clusters, one is associated with 1G stars thought to form from pristine gas with primordial helium (Y $\sim 0.25$). The second sequence is offset primarily along the \deltatwo\ axis due mostly to nitrogen enhancement tracked in the F336W filter and are thus enriched 2G stars expected to have helium abundances as high as Y $= 0.45$ (\deltay $= 0.2$), though upper limits are unknown. 

At fixed metallicity, offset of stars in the \deltaone\ axis can be attributed to helium enhancement, shown by \cite{Milone_2018} who used ATLAS12 and SYNTHE model atmospheres \citep{Kurucz_1970,Kurucz_1993,Sbordone_2004} to compute He, C, N, O and Mg simultaneously for each of the studied clusters. They find helium mass fraction enhancements of \deltay\ $= 0.01-0.10$ with the 2G populations being enhanced in helium relative to the 1G populations. The level of helium enhancement correlates with present day and initial cluster stellar mass \citep{Milone_2018,Gratton_2019}.

\omc\ has a uniquely complex set of subpopulations \citep{Lee_1999} making it a challenging cluster in which to infer helium enhancement. Among its stars there is a nearly 2.0 dex spread in metallicities, suggesting it is not a simple globular cluster but rather the surviving nuclear star cluster (NSC) of a stripped dwarf galaxy \citep[e.g.][]{Limberg_2022}. The formation of NSCs is still not fully understood but is expected to be a combination of globular cluster in-spiral as well as successive in-situ star formation events \citep{Neumayer_2020,Fahrion_2021}. A recently completed work provided the first comprehensive measurement of the age-metallicity relation in \omc, and found an unexpected two-stream feature with stars at intermediate metallicities having a bimodal age structure \citep{Clontz_2024}. However, because this age study was done with sub-giant branch stars, the correlation between these different populations and the abundance variations/chromosome diagram populations is not yet clear. Understanding the formation of \omc\ will provide insights into the phenomenon of multiple populations seen in complex clusters and galaxy centers. 

Because of this interest, there has been a rich history of both direct and indirect helium measurements in \omc. The determination of high helium abundances was first inferred from the HST photometric observations of multiple color-magnitude diagram sequences \citep{Bedin_2004,Norris_2004} and subsequent metallicity measurements \citep{Piotto_2005}, which strongly suggested significant helium enhancements (\deltay$\gtrsim 0.1$)\footnote{Note that many helium measurements are relative enhancements (\deltay) from an assumed primordial helium abundance mass fraction of $Y = 0.245$; we include a small metallicity dependence of $Y = 0.245 + 1.5\cdot Z$ via the isochrones used in this work}. \citet{King_2012} used deeper color-magnitude diagrams and improved stellar models to determine the helium abundance of the blue main sequence to be $Y = 0.39\pm{0.02}$ at an assumed metallicity of [Fe/H]$= -1.32$, while the red main sequence was fit with a primordial helium abundance and a more metal-poor isochrone ([Fe/H]$= -1.62$).  The helium abundances inferred by these papers singled out \omc{} as being uniquely enhanced in helium relative to both other globular clusters and field stars in the Milky Way \citep[e.g.][]{Piotto_2005}.

Spectroscopically, \cite{Dupree_2013} constrained helium abundances from direct fitting of the 1.08 $\mu$m chromospheric line for two low metallicity ($\rm{[Fe/H]} = -1.86$ and $-1.79$) \omc\ red-giant branch (RGB) stars. For their two stars they report $Y \leq 0.22$ and $Y = 0.39-0.44$, confirming a helium enhancement of at least \deltay $= 0.17$ among the stars. 
Other studies, including \cite{Hema_2020} and \cite{Reddy_2020} have shown that it is possible to constrain the helium abundance in individual stars by determining the discrepancy in Mg abundance measurements between the \ion{Mg}{1} and MgH lines. Hydrogen depletion affects the opacity and this decrement in the MgH line allows for a constraint on the hydrogen which is then converted to a helium abundance via a model assumption of the He/H ratio for the star. \cite{Hema_2020} find high helium enhancement in the two observed \omc\ red giants ($Y = 0.374$ and $Y = 0.445$). \cite{Reddy_2020} follow a similar procedure for 13 red giants and find a range of helium abundances with an \deltay $= 0.15\pm{0.04}$. The studies form the foundation of the spectroscopic evidence of significant enhancements in helium among \omc's subpopulations.

There have been a variety of approaches using photometric data for \omc{} to constrain helium. \cite{Joo_2013} generate synthetic color-magnitude diagrams, finding helium variations of \deltay $ = 0.16\pm{0.02}$ are needed to reproduce the features seen. \cite{Tailo_2016} perform a similar analysis adding a consideration of the C+N+O enhancement for the metal rich population and finding \deltay $= 0.12$ is needed for their models. \citet{Milone_2018} compare photometric measurements to synthetic spectra for 3084 metal-poor stars in \omc\ (consistent with those in our sample with $\rm{[Fe/H] \simeq -2.0}$) and find a helium enhancement between the most metal-poor 1G and 2G stars to be, on average, $0.033\pm{0.006}$ and a maximum \deltay~ $= 0.090\pm{0.010}$. A follow-up study by \cite{Milone_2020} follows a similar procedure, but this time divides the sample into 5 distinct clumps on the chromosome diagram. The population with the lowest \deltatwo\ color is considered the 1G population while the rest are considered 2G populations, labeled $\rm{A-D}$, each with increasing \deltatwo\ color. The maximum helium difference is found between the 1G and $2\rm{G}_{D}$ population and is \deltay $= 0.081\pm{0.007}$. These works form the basis for the photometric studies of the helium enhancement in \omc{} and suggest low helium enhancement in the metal-poor populations. 

With oMEGACat, a photometric and spectroscopic survey covering the full half-light radius of \omc, there is a new opportunity to study the helium enhancement in 2G stars. In this study we estimate the 1G to 2G helium differences as a function of metallicity for over 7,200 RGB stars, covering the full range of the stellar populations of this enigmatic cluster. Our new estimates use improved isochrone models, synthetic spectra, and individual abundance measurements from machine-learning techniques to derive more accurate helium constraints.

In Section \ref{sec:data} we present the source catalog and our sample selection. In Section \ref{sec:models} we describe the various models used in our measurements, in Section \ref{sec:calc_delta_y} we outline our methods, and in Section \ref{sec:results} we present our results. Discussion and conclusions are presented in Sections \ref{sec:discussion} and \ref{sec:conclusions}. 

\begin{figure*}
\includegraphics[width = \textwidth]{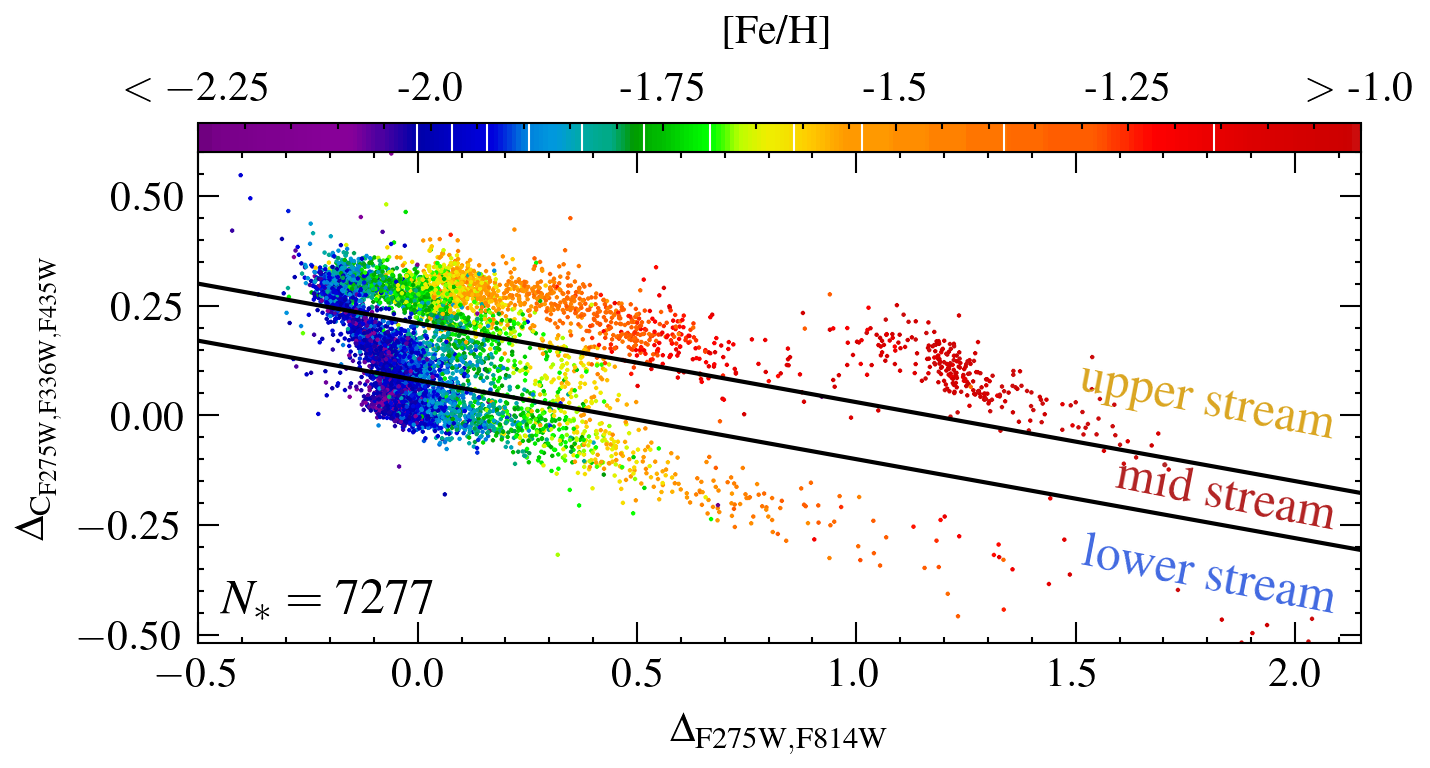}
\caption{\textbf{Chromosome/Pseudo-color diagram}: Our sample of 7,277 RGB stars in \omc\ is shown with each star colored by its metallicity. We separate the RGB stars separate into three distinct streams in this diagram using the diagonal black lines and color-coded labels. The edges of the [Fe/H] bins are indicated by white lines on the color bar (see also Table~\ref{table:results}) Spreads in the \deltaone\ and \deltatwo\ within metallicity bins are primarily due to light element abundance variations.}
\label{fig:chrom_map}
\end{figure*}

\section{Data} \label{sec:data}
For this study we use the oMEGACat catalog which contains information of individual stars of \omc{} from HST and MUSE observations. The photometric magnitudes are provided in 6 broadband HST filters, each having a correction for the differential reddening and instrumental effects \citep{Haeberle_2024a}. The metallicity is measured by full-spectrum fitting to each MUSE spectra and an atomic diffusion correction is applied \citep{Nitschai_2023}. More than 300,000 stars in oMEGACat have measurements in all photometric bands and metallicity information. We refer to the above two studies for detailed data reduction procedures. To match the metallicity to isochrone models used later, we convert $\rm{[M/H]}$ values to $\rm[Fe/H]$ using Equation 3 from \cite{Salaris_1993}:

\begin{equation}
    \rm{[Fe/H] = [M/H] - \log(0.638 \times 10^{[\alpha/Fe]} + 0.362)},
\end{equation}

assuming $\rm{[\alpha/Fe] = 0.3\,dex}$ \citep{Norris_1995, Johnson_2010}. This conversion has explicit dependence on the uncertainty in the abundance of each star, though this uncertainty is found to be similar to the typical metallicity error (0.08 dex). 

In this study, we use a subset of the RGB stars presented in \citet{Nitschai_2024}. To limit scatter in the chromosome diagram and avoid any impacts of dredge up while still retaining large numbers of stars with high spectroscopic S/N to ensure accurate metallicity and abundance measurements, we limit ourselves to stars with $\rm{14 \leq m_{F814W} \leq 17 mag}$. Using this magnitude range ensures consistent measurements of the helium abundance (Section~\ref{subsec:helium_ruler}). Also, the chromosome diagram of these stars clearly separates subsamples of stars due to the parallel sequences of these RGB stars on the color-magnitude diagrams. We use stars with F275W, F336W, F435W, and F814W measurements available from \citet{Haeberle_2024a} and cut using the HST quality flag. Our final sample contains 7,277 RGB stars. 

To obtain the pseudo-colors for the chromosome diagram, we use the fiducial lines from \cite{Nitschai_2024} to create the \deltaone\ \& \deltatwo\ colors. These fiducial lines vary with position on the color-magnitude diagram, but are applied in a consistent way to all data and models. We combine these data with the metallicities in \cite{Nitschai_2023} to produce the chromosome diagram in Figure \ref{fig:chrom_map}. This figure demonstrates that different groups in the chromosome diagram can be effectively separated by metallicity with low metallicities on the left hand side of the diagram, and higher metallicities to the right. Previous studies have divided the chromosome diagram into 1G and 2G stars using a single diagonal line \citep{Milone_2018} while other studies have further separated the 2G stars into two subpopulations \citep{Marino_2019}. Following the \citet{Marino_2019} work, we choose to divide our sample into three streams by utilizing the empirical 1G/2G separation, then we create a parallel line which separates the upper stream by cutting through under-dense regions in the chromosome diagram. The lower, mid, and upper streams contain 2071, 1920, and 3286 stars respectively. The stream separation is shown by the black lines in Figure \ref{fig:chrom_map}. We note that the lower/1G stars do follow the expectation of stars with varying metallicity and similar abundance as shown in the Appendix, and the slope of the line chosen is parallel to the expected variation in metallicity.

We further separate this sample into 12 bins based on metallicity. These are based on percentiles of the metallicity distribution similar to the selection in \citet{Clontz_2024}, but with smaller bins at higher metallicities to accommodate the complex structure visible in Fig.~\ref{fig:chrom_map}. The first 8 metallicity bins contain $\sim 727$ stars each (10\% of stars in each bin) with the last four containing 437, 364, 363, and 291 stars respectively (exact boundaries are given in Table~\ref{table:results}).

\section{Model Ingredients}\label{sec:models}
Spreads within metallicity bins in the chromosome diagram reflect the relative chemical abundances of the stars within that bin including variations in helium as well as other light element abundances. An accurate measurement of helium enhancements from the chromosome diagram requires a careful isolation and removal of effects from all other contributing abundance variations. We detail model ingredients required to make this measurement in this section.

To understand the impact of helium on the chromosome diagram, we use a purpose-built set of isochrones for \omc\ (Section~\ref{subsec:iso_models}). We account for other abundances' influence on the chromosome diagram using synthetic spectra models (Section~\ref{subsec:synthetic_spectra}). We also briefly discuss abundance estimates using our MUSE spectra in Section~\ref{subsec:abundance_estimates}. In Section~\ref{sec:calc_delta_y} we combine these model ingredients to show that changes in the \deltaone\ color in the RGB stars \omc\ are dominated by helium enhancements, and can be used to obtain accurate measurements of the helium enhancement after correcting for the impacts of other abundances.

\subsection{Isochrone Models} \label{subsec:iso_models}
The isochrones used in this study are built on the base model grids from the Dartmouth Stellar Isochrone Database \citep{Dotter_2008}. The improved isochrones tailored to \omc\ were first presented in \citet{Clontz_2024}, which contains more detailed descriptions and downloadable versions of the isochrones. These isochrones have an embedded C+N+O vs. [Fe/H] relation which was empirically constrained for RGB stars in \omc{} by \cite{Marino_2012}. We perform a linear fit to the data and for [Fe/H]$> -1.0$ we keep the C+N+O enhancement fixed. We also note that the same relation is found for 1G and 2G stars, according to their work. 

These updated isochrone models have several tuneable parameters. In this work we use models with $-2.5 < \rm{[Fe/H]} < -0.5$, and a fixed $\rm{[\alpha/Fe]}$ of 0.3 dex. We use two different helium abundances, one that we refer to as ``primordial" with $Y = 0.245 + 1.5\cdot Z$, and one that is similar to the highest helium abundances observed in \omc{}, $Y = 0.40$\footnote{Note that $Z$ is the mass fraction of elements heavier than helium and is calculated via the conversion from [Fe/H] given by 
$Z = Z_{\odot} \cdot 10^{[Fe/H]}$ where $Z_{\odot} = 0.014$ \citep{Asplund_2009}.}. 
Each isochrone is defined along a set of equivalent evolutionary points, covering the main sequence through the RGB. We use these models to constrain the chromosome diagram spread due to metallicity variations with a bin, described in Section \ref{subsubsec:feh_vector} and the helium ruler described in Section \ref{subsec:helium_ruler}. 

\subsection{Synthetic Spectra}
\label{subsec:synthetic_spectra}
While the isochrone models allow us to constrain the color differences on the chromosome diagram due to metallicity and helium abundance, they do not provide us with information about spreads due to contributions from other elemental abundance variations. Additionally, while the isochrones do incorporate a C+N+O relation with [Fe/H], they do not constrain contributions from C, N, and O individually, which are known to vary widely between subpopulations in \omc\ and other clusters \citep[e.g.][]{Marino_2012}. To quantify the impact of these light element abundance variations (other than helium), we turn to synthetic spectra. 

The synthetic spectral models generated are created following very closely the method used by \cite{Milone_2018}, using ATLAS12 and SYNTHE codes \citep{Castelli_2005, Kurucz_2005, Sbordone_2007}. Like \citet{Milone_2018} we consider the impact of C, N, O, and Mg, which have a significant impact on the chromosome diagrams. We simulate spectra using a surface and gravity appropriate to stars at the median F814W magnitude of our sample ($\simeq$ 15.5) and covering its full metallicity range. For each of these spectra, we then simulate spectra both with primordial (1G) abundances, and enhanced in one of the light element abundances based on measurements of 2G stars in \omc\ from \citet{Milone_2020}. Specifically, we simulate the following differences of enhanced$-$primordial abundances: $\Delta {\rm C} = -0.4$; $\Delta {\rm N} = +1.2$; $\Delta {\rm O} = -0.4$; and $\Delta {\rm Mg} = -0.4$. 

From the simulated spectra, we compute  the (F275W-F814W and $C_{275,336,435}$) colors of the stars both with primordial and enhanced abundances. These colors are then processed using the same fiducial lines used to produce our chromosome diagram. We use the differences between the primordial and enhanced abundance points on the chromosome diagram to create a reference vector that shows the impact of each element on the chromosome diagram position of a star. This procedure is repeated for each separate metallicity bin. The reference vectors are generated for all four elements (other than helium) expected to contribute to the chromosome diagram (C, N, O, and Mg). These reference vectors are then scaled based on the abundance variations measured between the streams  to arrive at the correction vector used in the \deltay calculation. Further discussion on the delta abundances used for each stream and each metallicity bin is given in Section~\ref{sec:calc_delta_y}. 

\subsection{Abundance Estimates}
\label{subsec:abundance_estimates}
Although the low spectral resolution of MUSE spectra (R of 1770 to 3590) makes it difficult to measure all chemical abundances, some can still be determined using neural network models like DD-Payne \citep{Xiang_2019}. 
This approach has already been successfully applied to other MUSE spectra and we refer to \cite{Wang_2022} for detailed procedures. 
A comprehensive analysis of DD-Payne-measured chemical abundances of \omc\ and their robustness and variations in chromosome diagram in a forthcoming paper (Wang et al, {\em in prep}). 
According to analysis by Wang et al.~({\em in prep}), the DD-Payne-derived Mg and O abundances in \omc, and their variations along the chromosome diagram are consistent with results from \cite{Milone_2020} and \cite{Johnson_2010}, respectively, and thus allow us to obtain individual stellar Mg and O abundance estimates for our full sample. 
The [Mg/Fe] abundances are obtained using DD-Payne model trained on APOGEE-Payne \citep{Ting_2019} for each of our stars with a median uncertainty of 0.05 dex. 
The [O/Fe] abundances use the model trained on GALAH DR2 \citep{Buder_2018},  with a median uncertainty of 0.14 dex. 
Two of the other abundances required for our corrections, nitrogen and carbon, are not available for individual stars. Wang et al. {\em in prep} are unable to constrain N abundances due to the lack of strong absorption lines in the MUSE spectral range, while they find significant systematics on the C abundance estimates. 
Therefore, in this work, we use DD-Payne-derived Mg and O abundances for our analysis. For carbon and nitrogen elements, we incorporate abundance constraints from previous studies on \omc{}, presented in detail in the next section. 
A more thorough examination of individual abundances apart from helium on the chromosome diagram will be presented in Wang et al. {\em in prep}. Here we use the Mg and O abundances solely to make the corrections for the abundances on the chromosome diagram in a metallicity dependent way.

\begin{figure*}
\centering
\includegraphics[width = \textwidth]{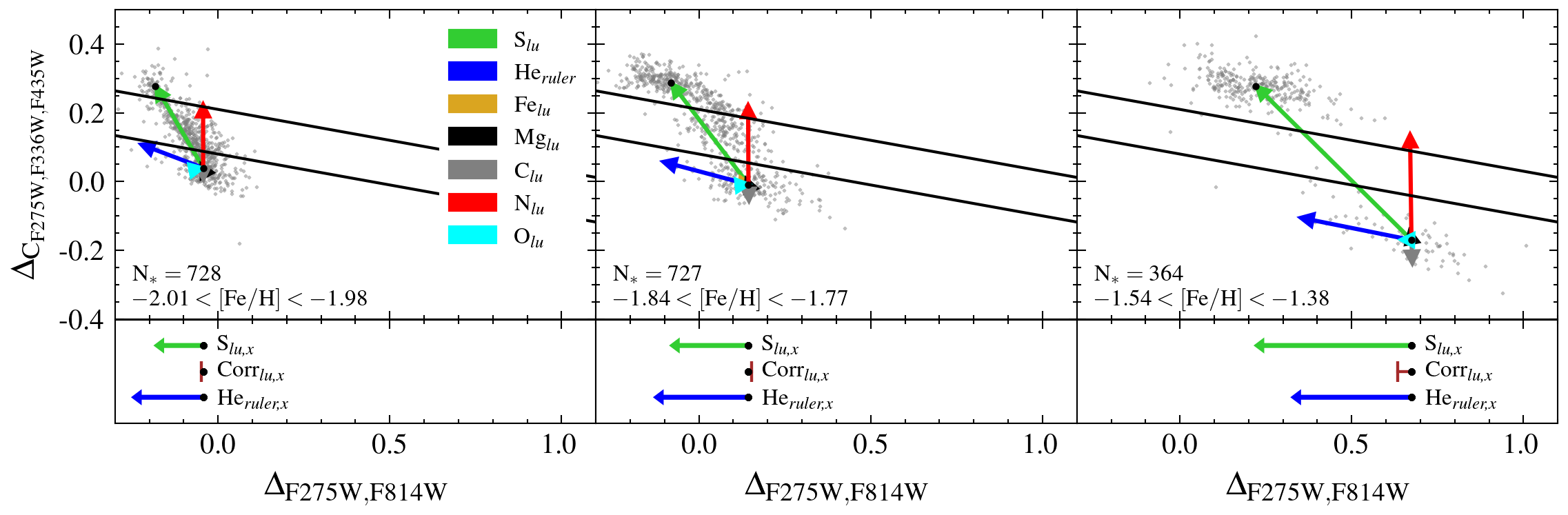}
\caption{\textbf{Determining Helium Differences using the Chromosome Diagram:} (\textit{Upper panels}) Stars in three different metallicity bins are shown in grey left to right. The star offset vector, the abundance correction vectors, and the helium ruler vector are all plotted in the upper panels extending from the lower stream centroid. (\textit{Lower panels}) The \deltaone\ components of each of the vectors are shown. The three black dots coincide with the lower stream centroid. The inferred \deltay\ is determined by scaling the helium ruler (blue vector) to the star offset (lime green vector) after corrections for the other abundances. Note that in each case, these corrections are small, and thus we attribute the bulk of the leftward offset in the \deltaone\ color between the upper stream stars and the lower stream stars to helium enhancement in the upper stream.}
\label{fig:method_vectors}
\end{figure*}

\section{Calculating Helium Enhancement}\label{sec:calc_delta_y}

For this study we aim to measure the helium enhancement between the streams of the chromosome diagram as a function of metallicity. We use the lower stream as a reference and measure the offset between it and the mid or upper streams to get the star offset vector (Section~\ref{subsec:star_offset_vector}). We then subtract the contributions expected from other elemental abundance variations from this star offset vector (Section~\ref{subsec:corrections}). Finally, we use a helium ruler (Section~\ref{subsec:helium_ruler}) to measure the remaining offset and scale this to a helium enhancement in each metallicity bin. 

We perform this calculation only in the \deltaone\ axis for two reasons. One, we do not have individual stellar constraints on nitrogen, which is known to dominate the contribution to the offset between streams in the \deltatwo\ axis \citep[e.g.][]{Milone_2020}. And two, the \deltaone\ of the chromosome diagram is constructed specifically to probe variations in metallicity and helium. Therefore, at fixed metallicity, we can more easily isolate helium enhancement. 

\subsection{Star Offset Vector}
\label{subsec:star_offset_vector}
For each of our metallicity bins described at the end of Section~\ref{sec:data}, we find the centroid of each stream on the chromosome diagram by taking the median \deltaone\ and \deltatwo\ of the stars occupying each. For the comparison between the upper and lower streams, for example, we then find the difference in each delta color and call this the star offset vector. Fig.~\ref{fig:method_vectors} shows this process for three metallicity bins covering the range of metallicities in the cluster. The x/\deltaone component of this vector is denoted as $S_{lu,x}$ and $S_{lm,x}$ between the lower and upper/middle streams respectively in Table~\ref{table:results}). The uncertainty on the offset vector length is found by taking half of the $1-\sigma$ range of values found when taking 100 bootstrap resamplings of the full RGB sample and recalculating its length. 

\subsection{Helium Ruler}
\label{subsec:helium_ruler}
Our measurement assumes the 1G stars are made from gas not enhanced in helium. To test if the variations in $\rm{[Fe/H]}$ can explain the offsets seen between the lower stream stars of neighboring bins we perform a similar analysis to that outlined in Sect. \ref{subsubsec:feh_vector}, this time using the difference in lower stream median metallicities between [Fe/H] bins. These vectors do an excellent job of describing the offsets seen between centroids in the x axis on the chromosome diagram for most of the metallicity range, further justifying our choice to separate the 1G and 2G populations with a straight diagonal line and suggesting that the lower stream primarily consists of stars with primordial helium abundance. However, in the last two metallicity bins lower stream star centroids sit down and to the left of the [Fe/H] vector location. One possible explanation is a spread in helium among the 1G population.

The top panels of Figure~\ref{fig:method_vectors} show the offsets between the primordial and helium enhanced isochrones as blue vectors in three of our metallicity bins; the primary offset in each panel is to the left along the \deltaone\ \ x-direction. We use the x-component of this vector as our helium ruler, which we notate as $\rm{He}_\textit{}{ruler,x}$ in Table~\ref{table:results} and Eq.~3 below (blue vector in bottom panels of Figure~\ref{fig:method_vectors}). 

\begin{figure*}[htb]
    \begin{minipage}[t]{.552\textwidth}
        \centering
        \includegraphics[width=\textwidth]{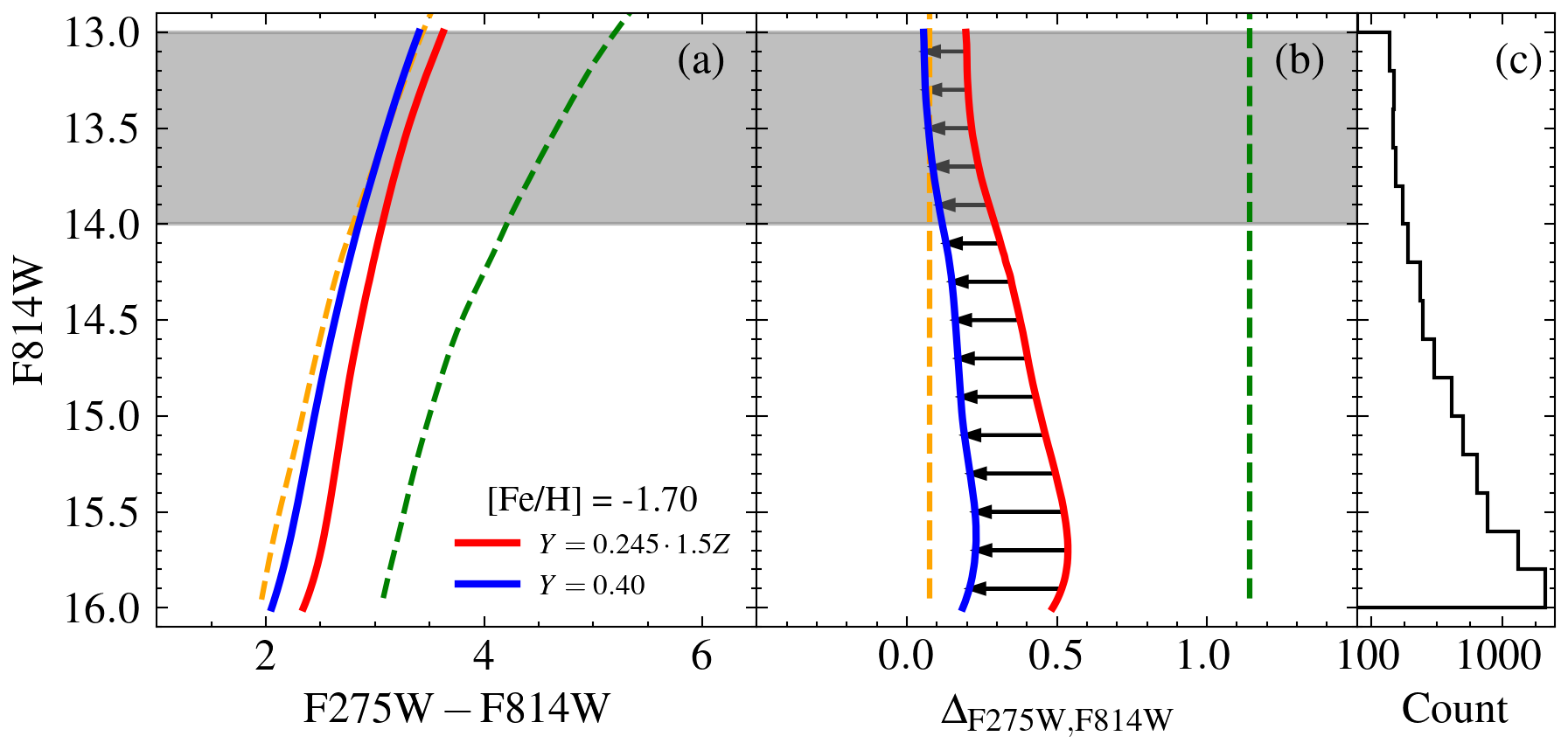}
    \end{minipage}
    \hfill
    \begin{minipage}[t]{.448\textwidth}
        \centering
        \includegraphics[width=\textwidth]{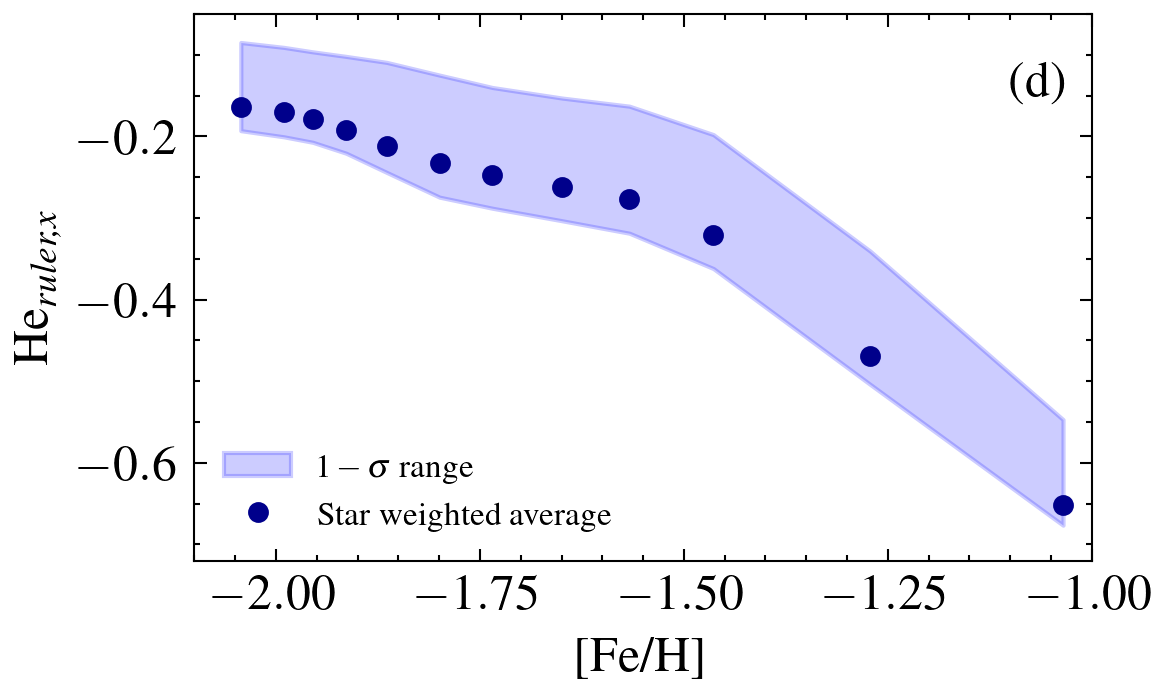}
    \end{minipage}  
    \label{fig:helium_ruler_creation_and_results}
    \caption{\textbf{Helium Ruler vs. [Fe/H]:}  {\em Left Panel --} These three figures show the derivation of the helium ruler. First on the far left, the color magnitude-diagram of two isochrones [Fe/H] $= -1.70$, but with primordial (red) and enhanced (blue) helium abundance are shown. The yellow and green lines show the fiducials used to verticalize the color-magnitude diagram and create the \deltaone\ color shown in panel (b). The black arrows in this panel show the distance between the verticalized isochrones in each magnitude bin. The histogram in panel (c) shows the number of stars at each magnitude that are used to weight the separation of the helium isochrones and obtain the helium ruler x-distance, He$_{ruler,x}$. The gray region at the top of the panel (b) shows magnitudes excluded from our analysis to ensure consistent length helium ruler vectors for all stars; the rulers at these magnitudes are clearly shorter. {\em Right Panel --} The derived He$_{ruler,x}$ in the \deltaone\ color as a function of metallicity. This corresponds to the expected offset in the \deltaone\/x-axis of the chromosome diagram for two populations with $\Delta Y \sim 0.15$. This difference is calculated at each metallicity across the full range of RGB magnitudes; the dark blue points show the He$_{ruler,x}$ determined from the weighted average of all the vectors in the left-middle panel. The blue band shows the 16th and 84th percentile of the vector lengths.}
\end{figure*}

To calculate the helium abundance differences between the streams we need to quantify the offsets in the chromosome diagram due to helium changes, which we call the helium ruler. We create a separate ruler for each of our 12 metallicity bins (see Section~\ref{sec:data} and Table~\ref{table:results}). For each, we take two isochrones with the median metallicity of a given bin, one with $Y = 0.245 + 1.5\cdot Z$ and one with $Y = 0.40$; Fig.~\ref{fig:helium_ruler_creation_and_results} panel (a) shows two isochrones at a single metallicity in the F275W$-$F814W color magnitude diagram. We interpolate each isochrone finely (steps of 0.005 mags) in and then verticalize the interpolated isochrones with the same fiducial lines used for generating the chromosome diagram  (Fig.~\ref{fig:helium_ruler_creation_and_results}, panel (b). We then find the median color difference of the projected points of the primordial and helium enhanced isochrone for each 0.2 magnitude wide bin. We combine these magnitude bin helium ruler measurements via a weighted average where the weight is the number of stars in each magnitude range (Fig.~\ref{fig:helium_ruler_creation_and_results}, panel (c). This combined measurement is the final helium ruler for a single metallicity bin. We repeat this process for each metallicity bin and plot the results in Fig.~\ref{fig:helium_ruler_creation_and_results} panel (d). The vector corresponds to the helium mass fraction difference of $\Delta Y_{ruler} = 0.40 - (0.245 + 1.5\cdot Z)$ and thus is $\sim$0.15 with a small variation between 0.1548 and 0.1529 at [Fe/H]$=-2$ and $-1$ respectively.
The He$_{ruler,x}$ values (which have no units, since they are differences in normalized colors) range from -0.177 at low metallicity and to -0.669 at higher metallicities.

We exclude brighter stars from our measurement due to shorter helium rulers at brighter magnitudes as shown in the grayed out region at the top of panel (b) in Fig.~\ref{fig:helium_ruler_creation_and_results}. We note that this effect is seen in the data as well; brighter RGB stars on the lower stream are offset to bluer \deltaone\ colors relative to the fainter stars, while those on the middle and upper stream align well; this provides further evidence that the effect we are seeing in the \deltaone\ color offsets are due to Helium abundance changes. The shaded region in Fig.~\ref{fig:helium_ruler_creation_and_results} shows the 16th and 84th percentile range of values for the \deltaone\ color difference across magnitude at a given metallicity. This range is roughly $\pm0.04$ at all metallicities and gives a sense of the potential systematic error in the measurement of our helium ruler. Because this range is nearly constant with metallicity this systematic uncertainty is larger at lower metallicities than at high metallicities. However, because we have weighted the ruler measurements based on the magnitude distribution of the data (Fig.~\ref{fig:helium_ruler_creation_and_results}, panels {(b)/(c)}), this spread in values is not equivalent to a 1$\sigma$ systematic error, and instead represents a worst-case scenario. We therefore do not include this potential systematic error in our quoted errors in our derived $\Delta Y$ values using our helium ruler.

Our helium enhancement measurements also make the assumption that the \deltaone\ offset scales linearly with \deltay. While we only have two helium abundances from our C+N+O vs. [Fe/H]-relation isochrones (Section~\ref{subsec:iso_models}), the original set of Dartmouth Isochrones \citep{Dotter_2008} provides three helium abundances with which we can test the linearity of this scaling. To do this we source the $Y = 0.245 + 1.5Z$, $Y = 0.33$, and $Y = 0.40$ isochrone models and verticalize them to obtain the delta colors.  
We calculate the helium ruler x-distance/\deltaone{} color difference between the $Y = 0.33$ and $Y = 0.245+1.5Z$ and the $Y = 0.40$ and $Y = 0.245+1.5Z$ using the same process as described above to create two helium rulers, one with \deltay$\sim 0.09$ and the other with $\sim$0.15.  We then compare how similar the \deltaone$/$\deltay{} of these two helium rulers are across the full range of metallicities we consider. The median fractional in \deltaone$/$\deltay{} between the two rulers is 0\% with a spread of 6.7\%. The difference is on the order of 0.6\% at [Fe/H] $\sim -1.7$ and varies with metallicity, with a maximum value 13.6\%.  The scatter in this relation is similar in size to the uncertainties we infer for the \deltay{} measurements.  We therefore conclude it is reasonable to consider the helium ruler x-component scales linearly with \deltay, and that the level of systematic error resulting from this assumption is similar to our 1$\sigma$ measurement errors.  

\subsection{Contribution from Other Elemental Variations}
\label{subsec:corrections}

\subsubsection{[Fe/H] Contribution}
\label{subsubsec:feh_vector}
New metallicity information for the full RGB from \citet{Nitschai_2023} enables the isolation of stars within a narrow range of metallicity on the chromosome diagram as shown for our 12 metallicity bins in Figs.~\ref{fig:method_vectors} and \ref{fig:primary_results}. However, some metallicity differences within a single bin's lower, mid and upper streams may remain and impact our helium calculation. To measure this impact, in each metallicity bin we calculate the difference in the median metallicity between the lower and mid streams as well as between the lower and upper streams. The error is the 1-$\sigma$ range found from bootstrapping our sample. For several metallicity bins there is no significant difference between the median metallicities of the streams. At the highest metallicities some offsets are seen, with  0.06 dex differences between the lower and mid streams in the two most metal-rich bins, attributed to small number of stars in each sample. 

To correct for any metallicity differences seen between streams, we take an isochrone with the median [Fe/H] of the lower stream, find the F275W-F814W color and the $C_{F275W,F336W,F435W}$ colors at $m_{F814W} = 15.5$ mag, and calculate their \deltaone\ and \deltatwo\ colors. We repeat this process for the isochrone with the median metallicity of the mid and upper stream stars. We then draw a vector from the lower-stream stars' isochrone chromosome diagram position to the mid and upper stream stars' isochrone chromosome diagram position to create our [Fe/H] vector for the comparison of these two streams in this metallicity bin. We use the x-component of these vectors ($\rm{[Fe/H]_{lm,x}}$ and $\rm{[Fe/H]_{lu,x}}$) in our calculation of \deltay. The contribution of the metallicity to the spread in the \deltaone\ axis is small and even zero for several metallicity bins. All values are given in Table \ref{table:abundances}.

\subsubsection{[Mg/Fe] Contribution}\label{subsubsec:mg_vector}
2G stars are expected to be depleted in Mg, as the Mg-Al chain converts Mg to Al. The previous helium determination of \citet{Milone_2018} focused only on the most-metal-poor populations and thus they assumed a fixed $\Delta [\rm{Mg/Fe}]$ between the 1G and 2G stars. To better constrain the $\Delta [\rm{Mg/Fe}]$ as a function of metallicity, we use abundance estimates derived from our MUSE spectra using the DD-Payne machine learning algorithm (see Section~\ref{subsec:abundance_estimates}). Using these values we calculate the median [Mg/Fe] for the stars in each stream within a given metallicity bin. Next, we find the $\Delta [\rm{Mg/Fe}]$ between the lower and upper stream as well as the lower and mid streams. We use synthetic spectra to calculate the contribution of the spread between the stream due to a fixed differences in Mg, then scale this contribution based on our measured contributions to get the Mg correction vectors ${\rm Mg}_{lm}$ and ${\rm Mg}_{lu}$. We use the x-components ${\rm Mg}_{lm,x}$ and ${\rm Mg}_{lu,x}$ in our \deltay\ calculations. The  contribution of the [Mg/Fe] abundance variations to the spread in the \deltaone\ axis varies with metallicity and all values are given in Table \ref{table:abundances}.
 
\subsubsection{[O/Fe] Contribution}\label{subsubsec:o_vector}
The contribution to the chromosome diagram shape attributed to oxygen is handled in a similar manner to Mg. Star-by-star oxygen abundances are derived from DD-Payne predictions. We calculate the median [O/Fe] abundance for each stream within each metallicity bin, then calculate the $\Delta [\rm{O/Fe}]$ between the streams. We follow the same process as with Mg, by modeling the color differences and scaling the reference vector by our measured abundance difference to get the $\rm{O}_\textit{lm}$ and $\rm{O}_\textit{lu}$ correction vectors. We use the x-components ($\rm{O}_\textit{lm,x}$ and $\rm{O}_\textit{lu,x}$) to calculate \deltay. The contribution of the [O/Fe] abundance variations to the spread in the \deltaone\ axis varies with metallicity and all values are given in Table \ref{table:abundances}.

\subsubsection{[C/Fe] Contribution}\label{subsubsec:c_vector}
No individual stellar carbon abundance are yet available for our data (Section~\ref{subsec:abundance_estimates}). Therefore, we adopt the $\Delta [\rm{C/Fe}]$ values given in Table 3 in \cite{Milone_2020} (which use C abundances from \citealt{Johnson_2010}). They separate their low metallicity stars into 5 groups. The 1G population is then compared with the four 2G populations ($\rm{2G_A-2G_D}$). The $\rm{2G_B}$ population has a \deltatwo\ color closest to our mid-stream of stars while the $\rm{2G_D}$ population corresponds best to our upper-stream of stars \citep[Fig.7 in][]{Milone_2020}. Therefore, we chose to adopt the $\Delta [\rm{C/Fe}]$ values from the comparisons between 1G and $\rm{2G_B}$ and $\rm{2G_D}$, $-0.2\pm{.09}$ and $-0.42\pm{0.08}$. We assume this value to be fixed as a function of metallicity. While this is certainly an oversimplification, we note the contribution to the correction from Carbon is quite small (see Table \ref{table:abundances}) and plausible variations between metallicities would not have a large enough impact on our helium measurements to change our results significantly. 

Following our previous modeling process for Mg and O, we produce models for the chromosome diagram color differences due this abundance variation. We scale our reference vector by the assumed abundance variation to get the $\rm{C}_\textit{lm}$ and $\rm{C}_\textit{lu}$ correction vector and use the x-components $\rm{C}_\textit{lm,x}$ and $\rm{C}_\textit{lu,x}$ in our \deltay\ calculation. The contribution of the [C/Fe] abundance variations to the spread in the \deltaone\ axis varies with metallicity and exact values can be found in Table \ref{table:abundances}.

\subsubsection{[N/Fe] Contribution}\label{subsubsec:n_vector}
As with carbon, we do not have individual stellar abundances available for nitrogen for stars in our sample. For this reason we choose again to adopt the $\Delta [N/Fe]$ values from the 1G/2G comparison in \cite{Milone_2020}. This gives us $\Delta \rm{[N/Fe]}_{lm} = 0.62$ and $\Delta \rm{[N/Fe]}_{lu} = 1.02$. We model this abundance variation as we do for other correction abundances, by scaling the reference vectors to obtain $N_{lm}$ and $N_{lu}$, then using the x-components of the resulting vectors ($N_{lm,x}$ and $N_{lu,x}$) in our \deltay\ calculations. The contribution of the [N/Fe] abundance variations to the spread in the \deltaone\ axis varies with metallicity and is given in Table \ref{table:abundances}.

\subsection{Helium Enhancement Calculation}
We now have all the ingredients required to calculate helium enhancements as a function of metallicity in \omc. We first take the x-component of a given metallicity bin's star offset vector $S_{lu,x}$ (Section~\ref{subsec:star_offset_vector}) and subtract off the x-components contributions from all relevant correction elemental abundances (Section~\ref{subsec:corrections}) to get the x-component of the remaining offset vector $R_{lu,x}$:

\begin{equation}\label{eq:resulting_vector}
   R_{lu,x} =  S_{lu,x} - Corr_{lu,x}
\end{equation}   
where:
\begin{equation}\label{eq:resulting_vector}
  Corr_{lu,x} = \rm{Fe}_\textit{lu,x} +  \rm{C}_\textit{lu,x} + \rm{N}_\textit{lu,x} + \rm{O}_\textit{lu,x} + \rm{Mg}_\textit{lu,x}
\end{equation}

These summed corrections, $Corr_{lu,x}$, are given in Table~\ref{table:results}, while the individual abundance corrections are given in Table~\ref{table:abundances}. Then, to calculate the helium abundance enhancement between the lower and upper streams, $\Delta Y_{lu}$, we scale the resulting vector by the helium ruler using the following equation: 

\begin{equation}
   \Delta Y_{lu} = {R}_{lu,x}  \cdot (\Delta Y_{ruler} / \rm{He}_\textit{ruler,x})
\end{equation}

To calculate \deltay\ between the lower and mid streams we follow the same prescription, denoting the relevant terms with the `lm' subscript. We repeat this for each metallicity bin. The error on these measurements is calculated as the error on the star offset vector (discussed in Section \ref{subsec:star_offset_vector}), added in quadrature with the uncertainties for each of the correction terms. The uncertainty in \deltay\ between the lower and mid-streams has a mean value of 0.011 and between the lower and upper streams the mean uncertainty is 0.009.

\section{Results} \label{sec:results}

\subsection{Helium Abundance Enhancement} \label{subsec:delta_y_results}

\begin{figure*}[h!]
\centering
\includegraphics[width = .9\textwidth]{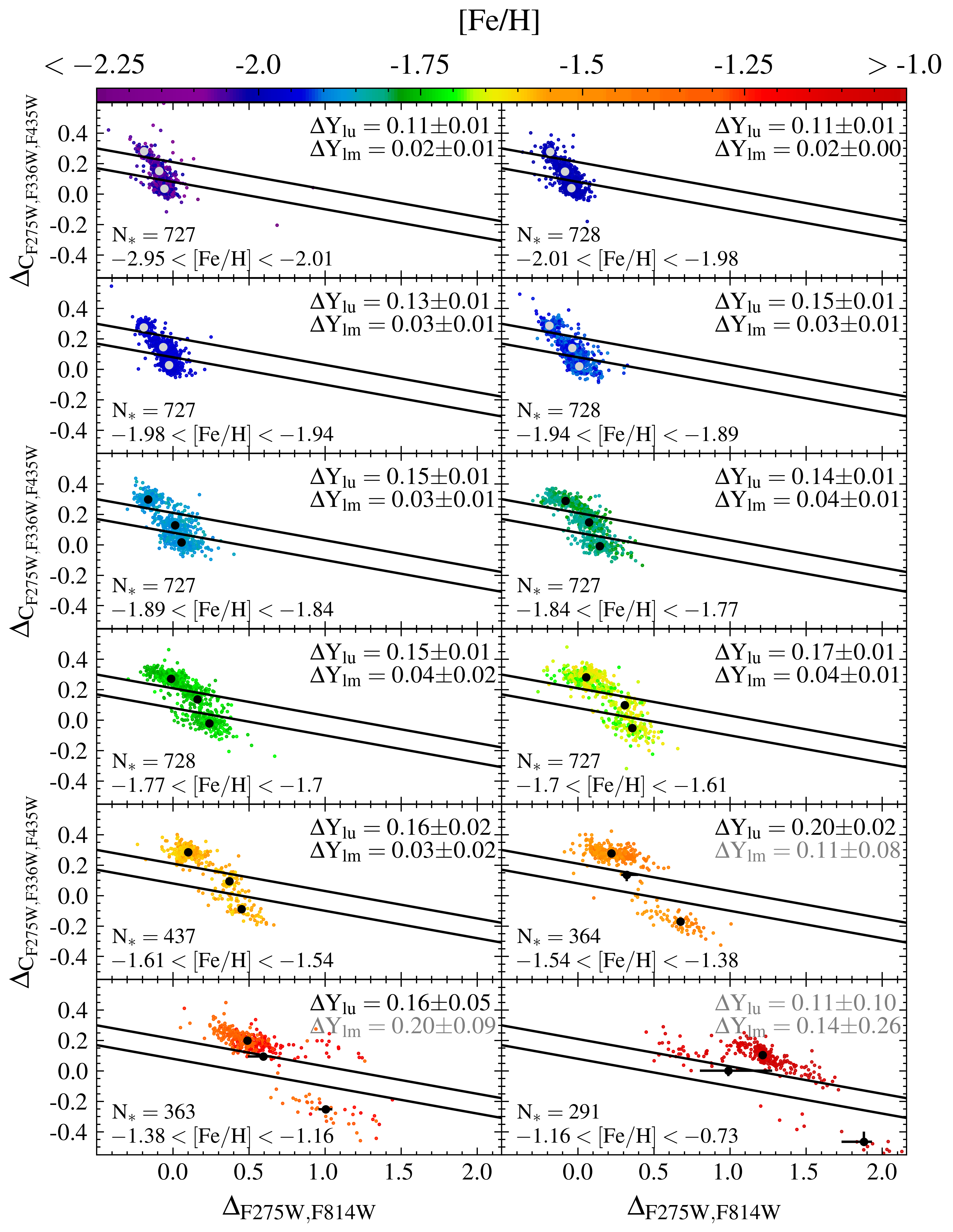}
\caption{\textbf{Deriving \deltay\ in twelve metallicity bins:} Within each panel we plot the stars belonging to a given metallicity bin along with the centroid markers for the lower, mid and upper streams (in gray or black to make them clearly visible). In the lower left corner of each panel we note the [Fe/H] range and the number of stars. In the upper right corner of each panel we report the $\Delta Y_{lm}$ and $\Delta Y_{lu}$ values with uncertainties, reporting highly uncertain measurements in gray font. The helium enhancement increases with metallicity.}
\label{fig:primary_results}
\end{figure*}

The results of our \deltay\ measurements are shown in Figure~\ref{fig:primary_results}.
Each metallicity bin is shown and can be read left to right top to bottom toward increasing metallicity. Within each panel we can see that the stream lines do isolate distinct clumps in the chromosome diagram across all metallicities. The chromosome diagram shows a similar spread in y-axis values across all bins, but the spread along the x-axis increases noticeably with metallicity, signaling an increase in helium enhancement. Also, the lower and middle streams are highly populated at the lower metallicities, but are almost non-existent at the highest metallicities while the upper stream remains well populated across all metallicities and includes most of the stars at the highest metallicities.

\begin{figure}[h]
\centering
\includegraphics[width = .5\textwidth]{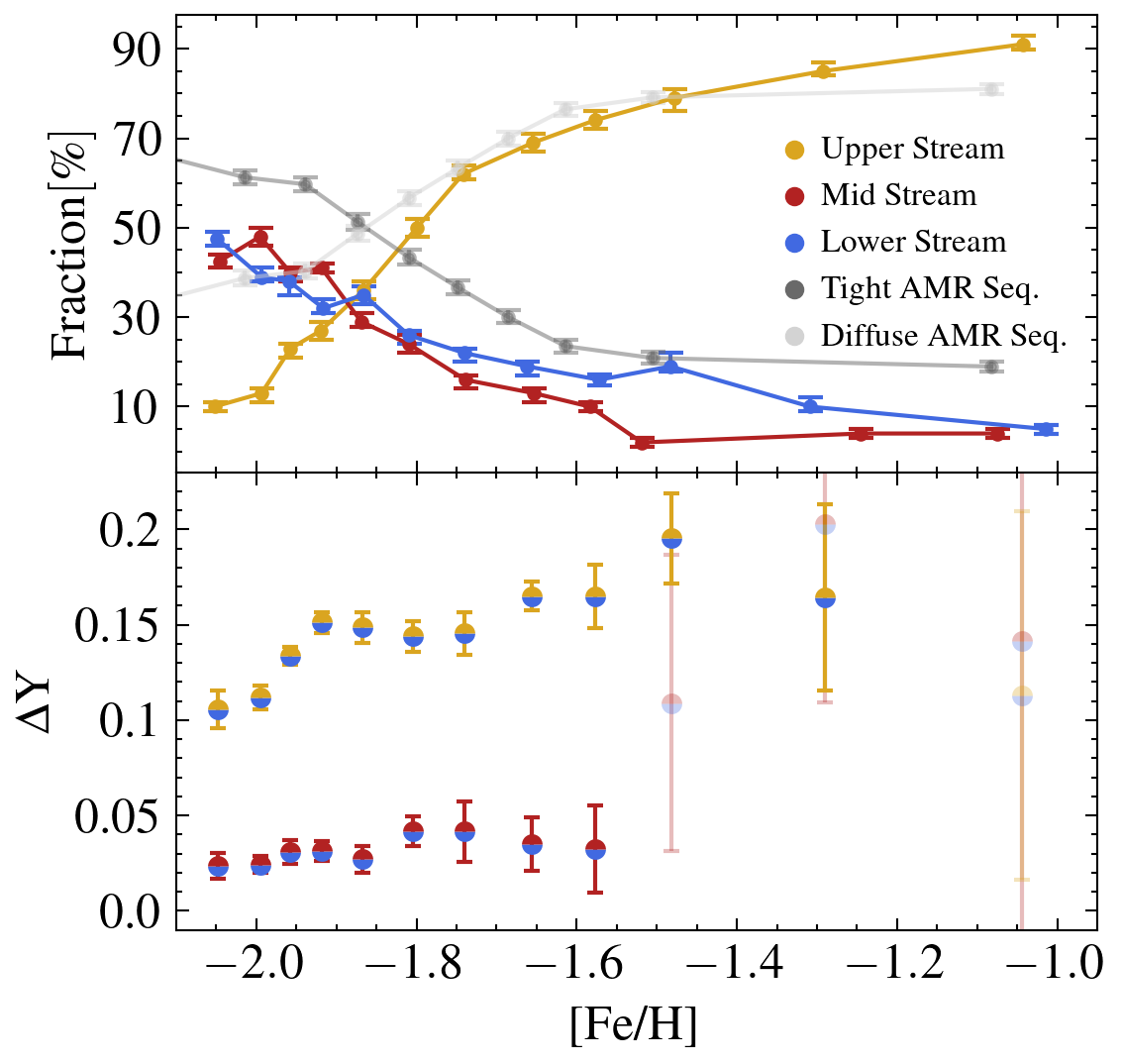}
\caption{\textbf{Helium Enhancement vs.~Metallicity:} (\textit{Top panel}) The fraction of stars in each of the three streams defined in Fig.~\ref{fig:chrom_map} as a function of metallicity (gold, blue, and red lines). Gray lines show the relative numbers of stars in the two different age-metallicity relationship tracks found by \citet{Clontz_2024}.
(\textit{Bottom panel}) \deltay\ for the lower-to-middle stream (blue/red points) and the lower-to-upper stream (blue/gold points) as a function of metallicity. The highest metallicity points are shown at low opacity because their results are very uncertain and are based on a small number of stars ($<$30).
}
\label{fig:streamwise_results}
\end{figure}

We summarize the results of our derived helium enhancements (\deltay) vs. metallicity bins in Table \ref{table:results} and the lower panel of Figure \ref{fig:streamwise_results}. Looking first at the lower vs. mid-stream star results we see helium enhancement of \deltay $ = 0.023\pm{0.007}$ at the lowest metallicity; stars at higher metallicities have slightly higher \deltay\ values, but remain $<$0.05 at all metallicities where they are well determined. 

For the lower vs. upper stream comparison we see at the lowest metallicity a helium enhancement of \deltay $= 0.11\pm{0.01}$ and a significant increase over the lowest 40\%ile of metallicities rising to \deltay $ = 0.15\pm{.005}$ at [Fe/H] = -1.92. The helium enhancement at higher metallicities is consistent with a flat line at $0.154\pm{0.004}$; the $\chi^2$ of a constant \deltay for the 9 higher metallicity bins is 1.07. A potential rise occurs above [Fe/H]$=-1.66$ with the highest bin having a value of $0.20\pm{0.02}$ at [Fe/H] of $-1.48$. The \deltay\ above this metallicity becomes increasingly uncertain due to the small number of lower stream measurements. Given an assumed primordial helium abundance ($Y \simeq 0.245$) for the lower stream, this suggests the upper stream stars at have a helium fraction of $Y \sim 0.40$ around the median metallicity of the cluster and up to $Y = 0.445\pm{0.02}$ at [Fe/H] of $-1.48$.

We also examine the number of enhanced stars as a function of metallicity by calculating the fraction of stars in the lower stream and comparing it to the fraction in the mid and upper streams for a given metallicity bin. 
The top panel of Figure \ref{fig:streamwise_results} shows the lower and mid streams constitute $48\pm{1}$\% and $43\pm{1}$ of the total stars each at low metallicities while the upper stream contains 9$\pm{1}$\%. The upper and midstream occupations then steadily drop to around 5$\pm{1}$\% each at the highest metallicities while the upper stream increases quickly up to [Fe/H] $\sim$ -1.74, then continues increasing more slowly up to a maximum of $>$90\% at the highest metallicities. The \deltay\ plateau in the upper stream (where it levels out at \deltay $\simeq0.15$) occurs at the metallicity where the all three streams have nearly equal numbers of stars.

\section{Discussion}\label{sec:discussion}
This work combines the techniques from many previous studies discussed above including isochrone models, synthetic spectra, and model stellar atmospheres with metallicity estimates and photometrically constructed chromosome diagrams to infer the helium enhancement of 2G stars in \omc. In this section, we discuss the implications of our primary results on the helium enhancements as a function of metallicity (Fig.~\ref{fig:streamwise_results}), comparing it to the literature results in $\omega$~Cen and beyond, and examining their implications for multiple stellar populations generally and $\omega$~Cen specifically.

\subsection{Comparison to Literature}\label{subsec:comparison_to_literature}
\citet{Dupree_2013} directly measured the helium abundances in two RGB stars with [Fe/H]$\simeq -1.8$ selected spectroscopically to be 1G and 2G stars based on their sodium and aluminum abundances. Based on the detection of the 1.08 $\mu$m line in one of the stars but not the other, they find a \deltay\ $\geq 0.17$. At a similar metallicity we find \deltay $= 0.15\pm{0.01}$, making these measurements consistent within 2$\sigma$. Their results suggest their two stars were members of the lower and upper streams. 
We also find consistency with other spectroscopic measurements, including \citet{Hema_2020} who measure  $Y = 0.374$ and $Y = 0.445$ (\deltay $= 0.13$ and \deltay $= 0.20$) for stars with [Fe/H] = -1.2 and -0.8 respectively. At similar metallicities we see a fixed value of \deltay $=0.154\pm{0.004}$ in the upper stream while a linear fit to our mid stream results predicts \deltay $=0.07\pm{0.04}$ and \deltay $=0.09\pm{0.06}$. \citet{Reddy_2020} sampled 20  stars across the metallicity range and report a \deltay $= 0.15\pm{0.04}$, consistent with the fixed enhancement seen in our upper stream stars. Overall, our results are consistent with the spectroscopic helium abundance measurements.

Comparing our results to the photometric study of metal poor poor stars $\rm{[Fe/H] \sim -2.0}$ by \cite{Milone_2018} who found a \deltay $= 0.033\pm{0.006}$, our constraints of the mid stream enhancement is \deltay\ value of $0.023 \pm{0.007}$, consistent with this value within the errors. Similarly the study by \cite{Milone_2020} which divided the 2G population into four, finds \deltay $= 0.016\pm{0.007}$ for their $2G_{B}$ population. Our upper stream has a \deltay\ value of $0.11\pm{0.006}$ and a centroid location (in chromosome diagram space) most comparable with their $2G_{D}$ population which has a \deltay $= 0.081\pm{0.007}$. The mid stream stars agree within $\rm{2\sigma}$ while our upper stream stars are significantly more enhanced. Differences in definitions of the populations, as well as the magnitude range used may account for these discrepancies despite similar methodology.

\begin{figure}[h]
\centering
\includegraphics[width = .5\textwidth]{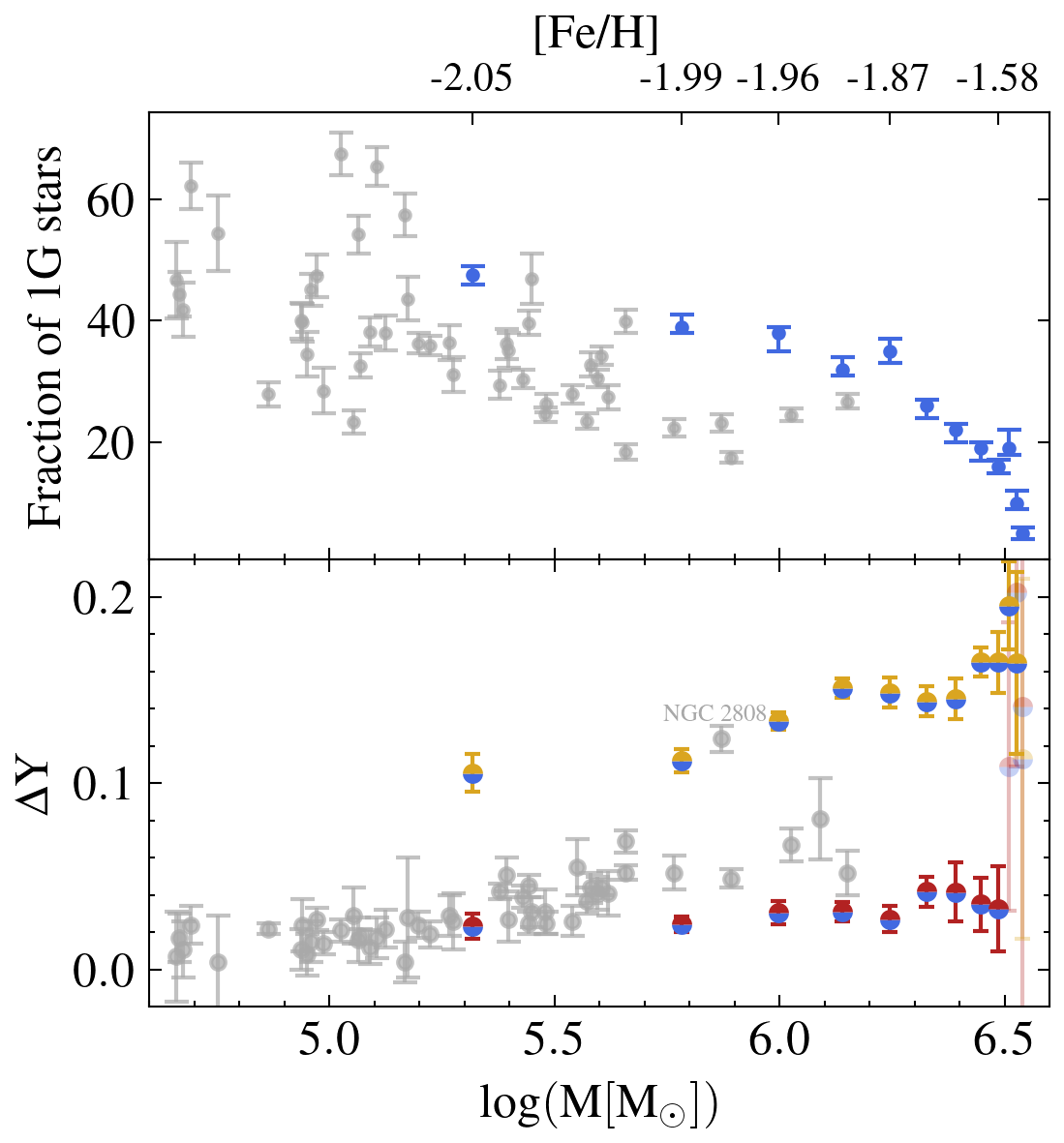}
\caption{\textbf{Trends with mass}: We compare the trends seen with globular cluster mass in fraction of 1G stars and helium enhancement to our results in \omc\ assuming that it built up its mass with metallicity. (\textit{Both panels}) In gray we plot a compiled globular cluster catalog using 1G fractions and \deltay values from \cite{Milone_2018} and masses from \cite{Baumgardt_2018}. \textit{Upper panel}: Both \omc\ and other globular clusters show a decreasing fraction of 1G stars, but the trend is more extreme in \omc{} than in other clusters. The trend at the highest masses (and thus metallicities) in \omc{} drops rapidly above [Fe/H]$= -1.86$ corresponding to an accumulated mass of $\rm{log(M_*) = 6.2}$. \textit{lower panel}: Our mid stream star helium enhancement follows a similar trend with mass to that of the GC catalog while our upper stream enhancement is much more rapid, only matched by NGC 2808.
}
\label{fig:delta_y_vs_mass}
\end{figure}

Lastly, we compare our stream fractions as a function of metallicity with the work of \cite{Bellini_2017c} where they use photometry to identify five primary main-sequence populations. The two dominant groups are the red main-sequences (rMS) and the blue main-sequence (bMS), thought to be the helium poor and helium rich populations respectively. We match their catalog with the spectroscopic catalog of \citet{Nitschai_2023} to assign metallicities to each their stars with subpopulation tags, giving us 7228 rMS and 6643 bMS stars for comparison. Similar to our lower and upper stream populations, which constitute 33\% and 40\% of our RGB sample each, while the rMS and bMS constitute 36\% and 33\% of the MS sample respectively. Plotting the fraction of rMS and bMS stars vs [Fe/H] we see a remarkably similar trend to that seen in the lower and upper streams of our RGB stars with the helium poor (rMS) population dominating below the median metallicity and the helium enhanced population (bMS) dominating at higher metallicities. This points to a direct connection between these groups; we plan to fully explore defining subpopulations spanning the MS and RGB in a future paper (Clontz et al., {\em in prep}).

\subsection{Comparison of \omc\ helium enrichment with other Millky Way GCs}
Here we examine how the helium and enrichment trends we see within a single cluster compare to the trends seen in the Milky Way GCs. First we note that the fraction of enriched stars has been found to strongly correlate with the present day and initial mass of the cluster they live in \citep{Milone_2017a,Gratton_2019}, with smaller 1G fractions at higher cluster masses. This trend is the same that we see in the fraction of lower stream (1G stars) in $\omega$~Cen {\em if} we consider that the metallicity tracks the mass build-up of the cluster, consistent with the age-metallicity relationship derived in \citet{Clontz_2024}. We show this direct comparison in the top panel of Fig.~\ref{fig:delta_y_vs_mass}. Specifically, the fraction of 1G stars in our lowest metallicity bin (corresponding to 10\% of the total mass in the cluster, or logM$\simeq$5.5) is $\simeq$45\%; within the scatter, but a bit higher than typical Milky Way clusters with similar mass.
The $<$10\% 1G fractions at the highest metallicites are lower than any other Milky Way cluster, consistent with $\omega$~Cen becoming the most massive cluster.

Along a similar vein, \citet{Milone_2018} find that the maximum \deltay\ in Milky Way clusters scales tightly with cluster mass just as we see a buildup in \deltay\ with increasing metallicity (and therefore presumably mass). The comparison of the Milky Way clusters helium enhancement with \omc~is shown in the lower panel of Fig.~\ref{fig:delta_y_vs_mass}. We find that at the lowest metallicity/mass end, the lower-to-middle stream \deltay\ is consistent with other Milky Way clusters clusters, while the lower-to-upper stream values are much higher. Apart from NGC~2808 \citep[which appears also to be affiliated with the Gaia-Enceladus-Sausage][]{Massari_2019}, the lower-to-upper stream \deltay\ values are higher than any Milky Way Clusters, while the Milky Way clusters fall somewhat above the lower-to-middle stream values. Like the Milky Way clusters, the lower-to-upper stream increases with increasing mass/metallicity.

Both these similarities could suggest that the trends we see with increasing metallicity in \omc\ could also trace an increase in cluster mass over time and that the mechanisms responsible for the trends seen in individual cluster formation were occurring over an extended period of \omc's assembly.

\subsection{Implication of Helium Enrichment Scenario}
As noted in many works \citep[e.g.][]{Norris_2004,Piotto_2005,Maeder_2006,Renzini_2008}, it is challenging to explain the origin of the extremely high helium abundances seen in \omc, and confirmed in this work. A review of polluter models and their predictions in context of Milky Way globular cluster abundance variations was made recently by \cite{Vaca_2024}. They find that no individual polluter model can explain the trends seen between globular clusters. With helium, they find that most abundance enhancements are not correlated with the production of helium with the exception of aluminum. The hottest hydrogen burning involves magnesium and aluminum, and thus we examine the correspondence between the helium abundance and Mg in \omc. The observed Mg depletion between the lower and upper stream (Table~\ref{table:abundances}) is nearly constant at $\sim -0.15$ dex except in the highest metallicity bins where the depletion between the two populations is lower. Thus there is no obvious correspondence between the increasing helium enrichment at low metallicities and any change in the Mg depletion.  Thus we do not find a similar result to \citet{Vaca_2024} that would tie together MgAl burning with helium production, although Mg is much less sensitive to abundance variations than Aluminium.

We note one additional pollution scenario that has not previously been invoked to explain globular cluster multiple populations: the growth and pollution from supermassive stars in an active galactic nucleus (AGN) accretion disk \citep{Cantiello_2021,Jermyn_2022}.  The recent confirmation of an intermediate-mass black hole at the center of the cluster \citep{Haeberle_2024b} suggests that star formation in \omc{} may have been accompanied by AGN accretion.  \citet{Jermyn_2022} find that accretion onto massive stars in the AGN disk could make these stars "immortal," with the stellar winds balancing the accretion rates for long periods of time.  These stars are expected to produce significant amounts of helium.  This scenario is akin to previous suggestions that supermassive stars could be one possible explanation for the multiple populations in globular clusters \citep[e.g.][]{Gieles_2018}; both these scenarios present the fascinating possibility of connecting the abundances of stars in \omc{} with the formation and accretion of its central black hole. 

\subsection{Implication of \omc{} formation mechanisms}
The trend we see of helium increasing linearly with metallicity over the lowest metallicities in the cluster argues in favor of a continuously enriching environment. This argues against a formation of the bulk of the cluster from individual clusters formed in separate environments as might be expected from the dynamical friction in-spiral formation mechanism of nuclear star cluster formation. Instead the increasing helium enrichment of the upper stream and the increase in the fraction of these stars suggests the polluter itself is increasing in efficiency over time as the mass of the cluster is built up, or that the increasing mass of the cluster leads to greater retention of pollutants. 

However, we know from the age-metallicity relation in \cite{Clontz_2024} that there exist at least two main formation channels for the cluster with different enrichment efficiencies.
We directly compare our enriched fraction results to the two-stream age-metallicity relation found in \cite{Clontz_2024} in the top panel of Fig.~\ref{fig:streamwise_results}. 
They found that the two age streams do not cleanly separate on a sub-giant branch chromosome diagram, although the correspondence between that diagram and the red giant branch one we present in Fig.~\ref{fig:chrom_map} is not straightforward.
It is interesting to note that the number of stars in the two streams as a function of metallicity follows a very similar trend to the fraction of stars we see in the different chromosome diagram streams (see the upper panel of Figure \ref{fig:streamwise_results}). In particular, the metallicity of the crossover point where the diffuse age-metallicity component starts to dominate is at similar metallicity ($\simeq -1.9$) to where the stars in the upper stream start to dominate the numbers. This might suggest a correlation between the diffuse age-metallicity component and the enriched population; ideally direct abundance measurements of the two components could resolve this issue.

\section{Conclusions} \label{sec:conclusions}
Helium is the second most abundant element in the Universe and is a useful tracer of second-generation stars in clusters with complex stellar populations. However, helium abundances in individual stars as well as stacked spectra have been notoriously difficult to constrain due to a number of observational and modeling challenges. Substantial progress has been made toward direct and indirect constraints for a subset of stars and have all pointed to significant helium enhancements in \omc. This study for the first time combines photometric and spectroscopic probes to infer the helium enhancement in \omc as a function of metallicity. The main findings of our analysis are as follows: 

\begin{itemize}
    \item In \omc, a spread in helium abundance ($\Delta Y \gtrsim 0.11$) is present at all metallicities. 
    
    \item For the upper stream stars, $\Delta Y$ in \omc~ increases with metallicity up to $\rm{[Fe/H]} \simeq -1.9$, then remains at a constant value of \deltay $= 0.154\pm{0.004}$. 
    
    \item The mid-stream stars have a relatively constant helium enhancement of $\simeq 0.030\pm{0.002}$. 
    
    \item The fraction of helium enhanced stars strongly increases with metallicity reaching 90\% at the highest metallicities. 

    \item The lower and upper streams exhibit similar trends in fraction of stars as a function of metallicity to the tight and diffuse age metallicity relation sequences in \cite{Clontz_2024} respectively. 
    
\end{itemize}

With this new information on the helium spread at fixed metallicity, we can begin to refine our models for fitting star formation histories to the observations of \omc's unique set of subpopulations. 

\begin{acknowledgments}

CC acknowledges the contributions to this work via the high performance computing resources at the University of Utah as well as the cluster computing resources of the Max-Planck Institute for Astronomy Heidelberg. ACS, ZW and CC acknowledge support from a Hubble Space Telescope grant GO-16777. M.A.C. acknowledges the support from FONDECYT Postdoctorado project No. 3230727. AFK acknowledges funding from the Austrian Science Fund (FWF) [grant DOI 10.55776/ESP542]. The authors thank Nicholas Stone for helpful discussions.

\end{acknowledgments}

%

\vspace{5mm}
\facilities{\textit{HST}(STScI), VLT:Yepun}


\software{\texttt{}}



\appendix

\begin{figure}
\centering
\includegraphics[width = \textwidth]{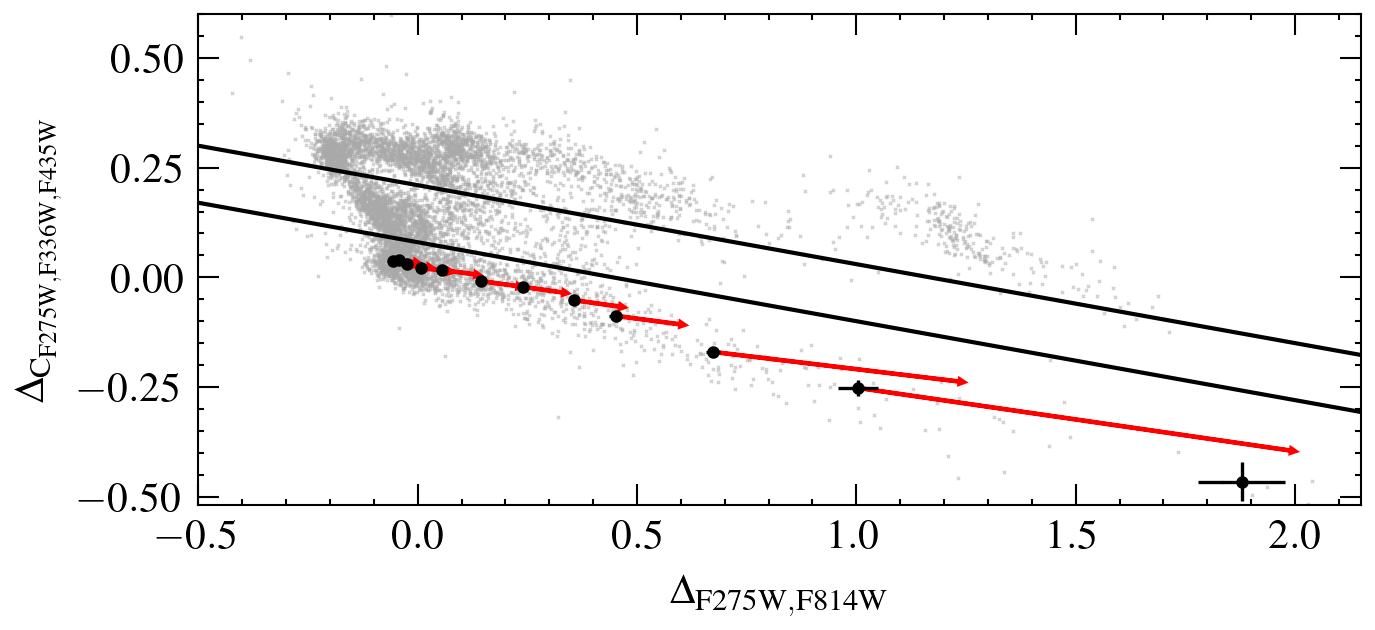}
\caption{\textbf{Lower Stream Metallicity Offsets}: Black points show the metallicity bin centroids for the lower stream and red arrows represent the spread on the chromosome diagram expected from the metallicity difference between the median metallicities of the lower stream stars in each [Fe/H] bin. These vectors are consistent with the offsets between the centroids suggesting that all lower stream stars have similar abundances. This justifies our assumption that these lower stream stars have primordial abundances. Note also that the vectors point relatively parallel to the streams; these inform our stream separtion lines shown in black. 
}
\label{fig:feh_vector_test}
\end{figure}

There are two fundamental assumptions in our analysis. The first is that the 1G stars can be separated from the 2G by a diagonal line that isolates the lowest stream on the chromosome diagram. The other is that the 1G consists only of stars with primordial helium abundance. Both of these can be tested using the same isochrone comparison and translation to the chromosome diagram that we use to create our helium ruler as detailed in Section~\ref{subsec:helium_ruler}, but using isochrones with varying metallicities rather than varying helium abundances. Specifically, we take the median metallicity of the lower stream stars from a given metallicity bin as well as the next most metal rich metallicity bin and calculate a vector which begins from the centroid of the lower metallicity lower stream centroid and has the magnitude and direction of the expected spread on the chromosome diagram due solely to the metallicity difference between that metallicity bin and the next highest metallicity one. These vectors are shown in red in Figure \ref{fig:feh_vector_test}. For most of the metallicity bins, the vector points from the given centroid almost directly to the next bin centroid, proving that the variation we see is consistent with a change only in overall metallicity.
For the last two metallicity bins we see the [Fe/H] vector over-predicts the spread between bins, which could be attributed to higher uncertainties in the metallicity estimates at these metallicities or it may be suggesting there is helium enhancement contributing to the movement of the higher metallicity bin centroids to smaller \deltaone\ values. Regardless, our helium enhancement measurements for these bins are already highly uncertain due to the low number of lower stream stars. 

\begin{sidewaystable}

\scriptsize
\centering
\begin{tabular}{ccccccccccccccccc} \\
\hline 
$[\rm{Fe/H}]_{min}$ & $[\rm{Fe/H}]_{med}$ & $[\rm{Fe/H}]_{max}$ & $\rm{He}_\textit{ruler,x}$ & $\sigma_{\rm{He}_\textit{ruler,x}}$ & $S_{lm,x}$ & $\sigma_{S_{lm,x}}$ & $\rm{Corr._{lm,x}}$ & $\sigma_{\rm{Corr._{lm,x}}}$ & $\Delta Y_{lm}$ & $\sigma_{\Delta Y_{lm}}$ & $S_{lu,x}$ & $\sigma_{S_{lu,x}}$ & $\rm{Corr._{lu,x}}$ & $\sigma_{\rm{Corr._{lu,x}}}$ & $\Delta Y_{lu}$ & $\sigma_{\Delta Y_{lu}}$\\
\hline
$-2.95$ & $-2.05$ & $-2.01$ & $-0.18$ & $0.04$ & $-0.033$ & $0.007$ & $-0.006$ & $0.005$ & $0.023$ & $0.007$ & $-0.132$ & $0.009$ & $-0.011$ & $0.008$ & $0.106$ & $0.010$ \\
$-2.01$ & $-1.99$ & $-1.98$ & $-0.18$ & $0.04$ & $-0.042$ & $0.004$ & $-0.013$ & $0.002$ & $0.024$ & $0.004$ & $-0.138$ & $0.005$ & $-0.005$ & $0.004$ & $0.112$ & $0.006$ \\
$-1.98$ & $-1.96$ & $-1.94$ & $-0.19$ & $0.04$ & $-0.038$ & $0.006$ & $-0.000$ & $0.003$ & $0.031$ & $0.006$ & $-0.167$ & $0.004$ & $-0.002$ & $0.003$ & $0.134$ & $0.005$ \\
$-1.94$ & $-1.92$ & $-1.89$ & $-0.20$ & $0.04$ & $-0.046$ & $0.005$ & $-0.005$ & $0.003$ & $0.031$ & $0.005$ & $-0.196$ & $0.005$ & $0.002$ & $0.003$ & $0.151$ & $0.005$ \\
$-1.89$ & $-1.87$ & $-1.84$ & $-0.22$ & $0.04$ & $-0.042$ & $0.006$ & $-0.003$ & $0.003$ & $0.027$ & $0.007$ & $-0.219$ & $0.007$ & $-0.003$ & $0.004$ & $0.149$ & $0.008$ \\
$-1.84$ & $-1.80$ & $-1.77$ & $-0.25$ & $0.04$ & $-0.071$ & $0.007$ & $-0.003$ & $0.004$ & $0.042$ & $0.008$ & $-0.224$ & $0.007$ & $0.009$ & $0.003$ & $0.144$ & $0.008$ \\
$-1.77$ & $-1.74$ & $-1.70$ & $-0.26$ & $0.05$ & $-0.080$ & $0.015$ & $-0.009$ & $0.005$ & $0.042$ & $0.016$ & $-0.251$ & $0.011$ & $-0.003$ & $0.004$ & $0.145$ & $0.011$ \\
$-1.70$ & $-1.66$ & $-1.61$ & $-0.28$ & $0.05$ & $-0.052$ & $0.013$ & $0.011$ & $0.008$ & $0.035$ & $0.014$ & $-0.304$ & $0.007$ & $-0.006$ & $0.006$ & $0.165$ & $0.008$ \\
$-1.61$ & $-1.58$ & $-1.54$ & $-0.29$ & $0.05$ & $-0.079$ & $0.021$ & $-0.017$ & $0.013$ & $0.033$ & $0.023$ & $-0.351$ & $0.015$ & $-0.037$ & $0.010$ & $0.165$ & $0.016$ \\
$-1.54$ & $-1.48$ & $-1.38$ & $-0.33$ & $0.05$ & $-0.353$ & $0.074$ & $-0.122$ & $0.024$ & $0.109\dag$ & $0.078$ & $-0.453$ & $0.022$ & $-0.040$ & $0.014$ & $0.195$ & $0.024$ \\
$-1.38$ & $-1.29$ & $-1.16$ & $-0.48$ & $0.04$ & $-0.447$ & $0.090$ & $0.179$  & $0.036$ & $0.203\dag$ & $0.094$ & $-0.516$ & $0.048$ & $-0.008$ & $0.017$ & $0.164\dag$ & $0.049$ \\
$-1.16$ & $-1.04$ & $-0.73$ & $-0.67$ & $0.04$ & $-0.853$ & $0.263$ & $-0.235$ & $0.035$ & $0.141\dag$ & $0.264$ & $-0.656$ & $0.096$ & $-0.161$ & $0.034$ & $0.113\dag$ & $0.097$ \\
\hline
\end{tabular}
\caption{\deltay~ Calculations: The metallicity bin edges and medians ($\rm{Fe/H}]_{min}$, $[\rm{Fe/H}]_{med}$, $[\rm{Fe/H}]_{max}$), the x-components of the helium ruler and error ($\rm{He}_\textit{ruler,x}$, $\sigma_{\rm{He}_\textit{ruler,x}}$; Section~\ref{subsec:helium_ruler}), the stars offset vector and error ($S_{lm,x}$, $\sigma_{S_{lm,x}}$; Section~\ref{subsec:star_offset_vector}), the total corrections vector x-component from all abundance differences ($\rm{Corr._{lm/u,x}}$, $\sigma_{\rm{Corr._{lm/u,x}}}$; Section~\ref{subsec:corrections}) for both the lower-mid stream comparsion (lm) and lower-upper stream comparison (lu), and the measured \deltay\ values and errors ($\Delta Y_{lm/u}$, $\sigma_{\Delta Y_{lm/u}}$).}
\label{table:results}
\end{sidewaystable}

\begin{table*}
\centering

\scriptsize
\begin{tabular}{ccccccccccccccc}
\toprule

\multicolumn{3}{c}{Sample} & \multicolumn{4}{c}{\textbf{[Fe/H]} (Nitschai et al. 2023)} & \multicolumn{4}{c}{\textbf{[O/Fe]} (Wang et al., \em{in prep.})} & \multicolumn{4}{c}{\textbf{[Mg/Fe]} (Wang et al., \em{in prep.})}\\

\cmidrule(rl){1-3} \cmidrule(rl){4-7} \cmidrule(rl){8-11}\cmidrule(rl){12-15} 
$[\rm{Fe/H}]_{min}$ & $[\rm{Fe/H}]_{med}$ & $[\rm{Fe/H}]_{max}$ & {$\rm{\delta_{lm}}$} & $\sigma_{\rm{\delta_{lm}}}$ &{$\rm{\delta_{lu}}$} & $\sigma_{\rm{\delta_{lu}}}$ & {$\rm{\delta_{lm}}$} & $\sigma_{\rm{\delta_{lm}}}$ &{$\rm{\delta_{lu}}$} & $\sigma_{\rm{\delta_{lu}}}$ & {$\rm{\delta_{lm}}$} & $\sigma_{\rm{\delta_{lm}}}$ &{$\rm{\delta_{lu}}$} & $\sigma_{\rm{\delta_{lu}}}$ \\
\midrule
$-2.95$ & $-2.05$ & $-2.01$ & $0.003$ & $0.004$ & $-0.004$ & $0.006$ & $-0.054$ & $0.024$ & $-0.067$ & $0.056$ & $-0.046$ & $0.009$ & $-0.149$ & $0.022$ \\
$-2.01$ & $-1.99$ & $-1.98$ & $-0.002$ & $0.001$ & $0.000$ & $0.002$ & $-0.001$ & $0.021$ & $0.005$ & $0.034$ & $-0.023$ & $0.006$ & $-0.140$ & $0.019$ \\
$-1.98$ & $-1.96$ & $-1.94$ & $0.001$ & $0.002$ & $0.002$ & $0.002$ & $0.008$ & $0.020$ & $0.034$ & $0.024$ & $-0.041$ & $0.009$ & $-0.149$ & $0.011$ \\
$-1.94$ & $-1.92$ & $-1.89$ & $-0.002$ & $0.002$ & $-0.003$ & $0.002$ & $-0.034$ & $0.020$ & $0.085$ & $0.019$ & $-0.040$ & $0.010$ & $-0.145$ & $0.007$ \\
$-1.89$ & $-1.87$ & $-1.84$ & $-0.003$ & $0.002$ & $-0.000$ & $0.002$ & $-0.014$ & $0.027$ & $0.039$ & $0.028$ & $-0.029$ & $0.011$ & $-0.150$ & $0.010$ \\
$-1.84$ & $-1.80$ & $-1.77$ & $0.000$ & $0.003$ & $0.010$ & $0.002$ & $-0.017$ & $0.027$ & $0.031$ & $0.026$ & $-0.030$ & $0.008$ & $-0.147$ & $0.007$ \\
$-1.77$ & $-1.74$ & $-1.70$ & $0.001$ & $0.003$ & $-0.002$ & $0.002$ & $-0.066$ & $0.029$ & $0.020$ & $0.022$ & $-0.025$ & $0.009$ & $-0.128$ & $0.008$ \\
$-1.70$ & $-1.66$ & $-1.61$ & $0.008$ & $0.006$ & $0.007$ & $0.004$ & $-0.001$ & $0.038$ & $-0.084$ & $0.027$ & $-0.010$ & $0.014$ & $-0.148$ & $0.008$ \\
$-1.61$ & $-1.58$ & $-1.54$ & $-0.012$ & $0.008$ & $-0.005$ & $0.007$ & $-0.029$ & $0.070$ & $-0.126$ & $0.051$ & $-0.010$ & $0.011$ & $-0.162$ & $0.014$ \\
$-1.54$ & $-1.48$ & $-1.38$ & $-0.036$ & $0.010$ & $0.005$ & $0.010$ & $-0.176$ & $0.130$ & $-0.202$ & $0.055$ & $-0.079$ & $0.060$ & $-0.154$ & $0.013$ \\
$-1.38$ & $-1.29$ & $-1.16$ & $0.063$ & $0.027$ & $0.016$ & $0.016$ & $-0.224$ & $0.119$ & $-0.323$ & $0.032$ & $-0.111$ & $0.028$ & $-0.111$ & $0.016$ \\
$-1.16$ & $-1.04$ & $-0.73$ & $-0.060$ & $0.033$ & $-0.028$ & $0.032$ & $-0.113$ & $0.051$ & $-0.195$ & $0.042$ & $-0.086$ & $0.035$ & $-0.054$ & $0.025$ \\
\bottomrule
\end{tabular}
\caption{\textbf{Metallicity Dependent Abundance Variations:} The metallicity bin edges and median, the difference in each elemental abundance between the streams ($\rm{\delta_{lm/u}}$ in dex), their relevant uncertainties constrained via bootstrapping ($\sigma_{\rm{\delta_{lm/u}}}$), as well as the reference for the source of each constraint are given. The correction vectors, which includes contributions from [O/Fe], [Mg/Fe], [C/Fe], and [N/Fe] are given in Table \ref{table:results}. See Section \ref{subsubsec:c_vector} and Section \ref{subsubsec:n_vector} for constraints on [C/Fe] and [N/Fe].}
\label{table:abundances}
\end{table*}

\bibliography{main_bib}
\bibliographystyle{aasjournal}


\end{CJK*}
\end{document}